Al doped graphene: A promising material for hydrogen storage at room temperature

Z. M. Ao,[†,‡] Q. Jiang,[‡,*] R. Q. Zhang,[‡] T. T. Tan,[†] S. Li[†,*]

School of Materials Science and Engineering, The University of New South Wales, NSW 2052, Australia and Key Laboratory of Automobile Materials, Ministry of Education, and Department of Materials Science and Engineering, Jilin University, Changchun 130025, China

**Abstract:**

A promising material for hydrogen storage at room temperature−Al doped graphene was proposed theoretically by using density functional theory calculation. Hydrogen storage capacity of 5.13 wt% was predicted at $T = 300$ K and $P = 0.1$ Gpa with adsorption energy $E_b = -0.260$ eV/H$_2$. This is close to the target of 6 wt% and satisfies the requirement of immobilization hydrogen with $E_b$ of -0.2 ~ -0.4 eV/H$_2$ at ambient temperature and modest pressure for commercial applications specified by U.S. Department of Energy. It is believed that the doped Al varies the electronic structures of both C and H$_2$. The bands of H$_2$ overlapping with those of Al and C synchronously are the underlying mechanism of hydrogen storage capacity enhancement.

[*] To whom correspondence should be addressed. e-mail: jiangq@jlu.edu.cn (Q. Jiang), e-mail: sean.li@unsw.edu.au (S. Li).
[†] The University of New South Wales
[‡] Jilin University

**I. INTRODUCTION**

In recent years, hydrogen-based fuel systems have been considered to be a highly important topic of research for future energy schemes as hydrogen is a more efficient fuel in comparison to the existing carbonaceous fossil fuels. [1,2] Despite many recent technological developments in the hydrogen-based fuel systems, it is still an enormous challenge to have a safe and efficient reversible hydrogen storage systems at ambient conditions.[2] One of the possible ways for such storage is the efficient hydrogen adsorption/desorption in a controllable system. Carbon based materials are candidates for such a purpose. Although several mechanisms of hydrogen storage through both physisorption and chemisorption have been proposed,[3-7] most of these efforts are far to reach the target of 6 wt% and immobilization hydrogen with binding strength of -0.2 ~ -0.4 eV/$H_2$ at ambient temperature and modest pressure for commercial applications specified by U.S. Department of Energy (DOE).

With density functional theory (DFT) simulations, it was predicted that a single ethylene molecule can form a stable complex with two transition metals, thus adsorbing ten $H_2$ molecules and lead to a high storage capacity of ~14 wt%.[8] In addition, the highest $H_2$ storage capacity of 13 wt% in a fullerene cage with twelve Li atoms capped onto the pentagonal faces was calculated.[9] This system has average adsorption energy $E_b$ = -0.075 eV/$H_2$. However, all the DFT results are in the ideal condition at the temperature of $T = 0$ K, their performances at the DOE specified operation conditions are unclear.

Since carbon nanostructures have high surface areas, and thermal stability along with unique mechanical properties, improvement of their adsorption capacity by suitable modification would be of immense interest.[3-9] Thus, the storage of hydrogen molecules by carbon nanostructures is still an important issue and deserves more attention. For example, the potential of graphene as hydrogen storage materials through doping is investigated. The advantages of graphene are: (1) a large surface for hydrogen adsorption, (2) economical and scalable production,[10] and (3) the strongest material ever measured.[11]

On the other hand, the prospect of $AlH_3$ and related aluminum hydrides as hydrogen storage materials have recently become the focus of renewed interest [12,13] as their potentially large hydrogen capacity of ~10 wt%. These materials are thermodynamically unstable in ambient, but it is kinetically stable without much lost of hydrogen for years. However, extremely high hydrogen pressure (exceeding 2.5 GPa) is required to load up the hydrogen. While these hydrides possess a small negative enthalpy of formation,[13] for practical applications there remains the critical issue of a large hydrogen removal energy that hinders the H-desorption process. The origin of this energy barrier lies in the rather strong mixed ionic and covalent bonds [13] formed between Al and H. In order for $AlH_3$ to become a feasible H-storage material, it is essential to develop a technique to significantly reduce the hydrogen removal energy and thus enable a controllable kinetics for hydrogen desorption.

However, there is another way for metallic atoms to storage hydrogen, which is further decreasing the interaction between Al and H. In this way, the weak chemisorption can be changed into strong physisorption. For physisorption of hydrogen storage, it requires a strong interaction between $H_2$ molecule and the materials on the surface with an area as large as possible. These superior characteristics of graphene and Al for hydrogen storage lead to an attempt of Al doped graphene to consolidate these advantages and to see whether graphene can take the role to weaken the interaction between Al and H or Al can take the role to enhance the adsorption of $H_2$ on graphene. In this work, the adsorption behavior of $H_2$ in Al doped graphene was studied by DFT calculation. In addition, we processed the *ab initio* molecular dynamics (MD) calculation to investigate the effects of temperature and pressure on the corresponding adsorption and desorption abilities of this system on hydrogen storage system.

## II. CALCULATION METHODOLOGY

All DFT calculations are performed in Dmol[3] code.[14] Previous studies [15,16] have shown that the local density approximation (LDA) prediction of the physisorption energies of $H_2$ on the surface of graphite and carbon nanotubes are in good agreement with experiments. The reliability of LDA can be ascribed to the following facts:[15] (1) When the electron densities of $H_2$ and graphene overlap weakly, the nonlinearity of the exchange-correlation energy density functional produces an attractive interaction even in the absence of electron density redistribution; (2) The overestimated binding energy by LDA [17,18] may compensate for the insufficient account of van der Waals interactions.[15] In contrast, DFT calculation using a uniform generalized gradient approximation (GGA) produced a purely repulsive interaction. Using a GGA-PW91 functional, a repulsive interaction between $H_2$ and a graphene layer and also between $H_2$ and a (6,6) carbon nanotube was obtained.[19] This contradicts the experimental findings.[20] It was noted that LDA calculations well reproduce the empirical interaction potentials between graphitic layers and also in the other graphitic systems for distances near to the equilibrium separation although the LDA is not able to reproduce the long-range dispersion interaction.[21] Therefore, LDA is used to implement the task of this work. To ensure the calculated results being comparable, the identical conditions are employed for the isolated $H_2$ molecules and the graphene, and also the adsorbed graphene system. The $k$-point is set to 6×6×2 for all slabs, which brings out the convergence tolerance of energy of 1.0×10$^{-5}$ hartree (1 hartree = 27.2114 eV), and that of maximum force is 0.002 hartree/Å.

In the simulation, three-dimensional periodic boundary condition is taken and H–H bond length is set to $l_{H-H}$ = 0.74 Å, which is consistent with the experimental results.[22] The graphene used in our simulation consist of a single layer of 3×3 supercell with a vacuum width of 12 Å to minimize the interlayer interaction. Further increasing of the vacuum width has only negligible consequence on the simulated results but is great at the expense of computation. All atoms are allowed to relax in all energy calculations. The adsorption energy $E_b$ between the $H_2$ gas molecule and graphene is defined as,

$$E_b = E_{H2+graphene} - (E_{graphene} + E_{H2}) \qquad (1)$$

where the subscripts $H_2$+graphene, graphene, and $H_2$ denote the adsorbed system, isolated graphene and $H_2$ molecules, respectively.

For the Al doped graphene, the concentration of Al is 12.5 at% with the additional constraining that there is only one Al atom per graphene hexagonal ring (Fig. 1) to avoid the doped metal Al atoms clustering on graphene.[23] For $H_2$ adsorption on the Al doped graphene, there are 4 top sites of T1, T2, T3 and T4, and 3 bridge sites of B1, B2 and B3, and 2 center sites of C1 and C2, as shown in Fig. 1. (In this figure, a larger simulation cell is given in order to better display the different adsorption sites on the Al doped graphene. In actual simulation, the simulation cell is taken as large as Fig. 2.) At each adsorption site, there are two highly symmetrical adsorption configurations, namely $H_2$ molecule resides parallel or perpendicular to the graphene surface. Therefore, together 18 adsorption configurations for $H_2$ adsorbed on the Al doped graphene are present.

Due to the periodicity of $H_2$ adsorbed in intrinsic graphene or Al doped graphene systems, we have selected a unit cell with eight C atoms and one $H_2$ or seven C atoms, one Al atom and one $H_2$ (see Fig. 2). If we place a $H_2$ at any location of the cell, the distance from this $H_2$ to other $H_2$ molecules in the nearest cells is 4.920 Å. This large separation, compared to the bond length of $H_2$ (0.740 Å), would ensure that there is no interaction between $H_2$ molecules in the different cells.[24]

To measure $H_2$ adsorption capability of Al doped graphene at room temperature and modest pressure, we performed *ab initio* MD calculation with CASTEP (Cambridge Sequential Total Energy Package) code based on the structure obtained by DFT above, which utilizes plane-wave pseudopotential to perform first principle quantum mechanics calculations.[25] LDA with the Ceperley-Alder-Perdew-Zunger (CAPZ) function [26,27] was employed as exchange-correlation functions, cutoff energy $E_c$ = 280 eV and $k$-points is 6×6×2. Each MD simulation was performed in *NPT* statistical ensemble, i.e. constant numbers of atoms *N*, pressure *P* and *T*, with *T* = 300 K and *P* = 0.0001~1 GPa. Time step of 1 *fs* was selected and simulation time *t* at a particular *T* was 2.5 *ps* where the total energy fluctuation was in the range of 0.01%. The same *t* was selected for $H_2S$ dissociation on the Fe(110) surface.[28] A Verlet algorithm [29] was used to integrate the equations of motion, with *T* controlled by algorithm of Nose,[30] and *P* was controlled according to the Parrinello-Raham algorithm.[31]

### III. RESULTS AND DISCUSSION

After geometry relaxation, $E_b$ values and the corresponding structure parameters of the 18 adsorption configurations for $H_2$ adsorbed in the intrinsic graphene are listed in

Table 1. It was found that the most favorable configuration is $H_2$ adsorbed on the center site of the carbon ring with $E_b$ = -0.159 eV as shown in Fig. 2(a) and the distance between $H_2$ and the graphene $d$ = 2.635 Å. The results are consistent with other reported results of $E_b$ = -0.133 eV and $d \approx 2.8$ Å.[16] The small magnitude of $E_b$ shows that the system is in the weak physisorption region. It indicates that the intrinsic graphene is not suitable for hydrogen storage.

For the adsorption of $H_2$ in the Al doped graphene, the corresponding results are also listed in Table 1. In light of Table 1, the most favorable position with $E_b$ = -0.427 eV for the $H_2$ molecule is shown in Fig. 2(b). The distance between $H_2$ and the doped Al, $l$ = 2.083 Å while that between $H_2$ and carbon layer, $d$ = 2.073 Å. As seen from Table 1, the interaction reaches the strongest when both $l$ and $d$ are minimized. The adsorption of $H_2$ in the Al doped graphene is much larger than that in other systems, such as $E_b$ = -0.41 eV/$H_2$ in Ti-$C_2H_4$-graphene system,[8] and $E_b$ = -0.08 eV/$H_2$ in 12-Li-doped fullerene.[9] However, it still belongs to the physisorption system as the long distance between the doped graphene and the adsorbed $H_2$. Therefore, this strong physisorption interaction would be ideal for hydrogen storage, which can adsorb more $H_2$ molecules.

To understand the enhancement effect of the doped Al on the $H_2$ adsorption, the projected electronic density of states (PDOS) of the adsorbed $H_2$, the doped Al and the C atoms in both $H_2$/graphene and $H_2$/Al-doped-graphene systems are plotted and shown in Fig. 3. Fig. 3(a) shows the PDOS of $H_2$/graphene system. The main peaks of $H_2$ are located at –4.37 eV and 6.92 eV. However, the main peaks of intrinsic graphene are located between 9 and 13 eV. Therefore, the interaction between $H_2$ molecule and the intrinsic graphene is very weak because of non-overlapping of electrons in these substances, where $E_b$ is small. On the other hand, for the $H_2$/Al-doped-graphene system shown in Fig. 3(b), the main peaks of $H_2$ are located at –8.15 eV, 5.74 eV, 6.52 eV, and 7.51 eV, respectively. The bands of $H_2$ interact with both the doped Al and the C atoms synchronously at the positions indicated by the dash lines, showing a strong interaction between $H_2$ and the Al doped graphene where $E_b$ is the largest. In addition, the doped Al changes the electronic structures of both $H_2$ and the graphene, and both their PDOSs shift towards the lower energy. It exhibits that the $H_2$/Al-doped-graphene configuration is a much more stable system.

Table 2 shows the charge distribution in both the $H_2$/graphene and $H_2$/Al-doped-graphene systems using Mulliken analysis. Before and after $H_2$ adsorption, the charge variation for the former is little while it is significant for the latter. In addition, H6 has much more positive charge than H5. Thus, the interaction between $H_2$ and the Al doped graphene is mainly achieved through H6. The interaction between the band at the location of the highest peak of PDOS plot of $H_2$ and that of C atoms implies a strong interaction between the $H_2$ and C atoms, as shown in Fig. 3(b).

The illustrations of electron density distribution for the $H_2$/graphene and $H_2$/Al-doped-graphene systems are shown in Fig. 4. In the system of $H_2$/graphene [Fig. 4(a)], no electron exists in the region between $H_2$ and C layer while some electrons appear in the region among $H_2$, Al atom and C layer in the system of $H_2$/Al-doped-graphene [Fig. 4(b)]. This supports the notion that the $H_2$/Al-doped-graphene possesses a much stronger $H_2$ adsorption ability.

After understanding the mechanism of the enhancement for $H_2$ adsorption in the Al doped graphene, it is important to determine how much $H_2$ molecules can be adsorbed on the 3×3 layer surface. We constructed an adsorption configuration with 3 $H_2$ molecules adsorbed in the three favorable C1 adsorption positions on the topside of the doped system. After geometry relaxation, the atomic structure is shown in Fig. 5(a). It has $E_b$ = -0.303 eV/$H_2$, which satisfies the requirement of $E_b$ = -0.20 ~ -0.40 eV/$H_2$ at room temperature [3-6] set by DOE although the value of 5.1 wt% of $H_2$ adsorbed is slightly below the DOE′s 6 wt% target.

In order to understand the effect of the adsorbed $H_2$ molecule number on the $E_b$, the configuration with 6 $H_2$ molecules adsorbed in the Al doped graphene in the favorable C1 adsorption positions on both sides was calculated. It is found that $E_b$ = -0.164 eV/$H_2$, which is much smaller than the $E_b$ for the above case where the Al doped graphene adsorbed 3 $H_2$ on one side of graphene. In addition, the adsorption with 8 $H_2$ molecules in the Al doped graphene was also calculated, and it is found 2 $H_2$ molecules were released. In the other words, the interaction between $H_2$ molecules would weaken the adsorption and the saturated number of $H_2$ molecules adsorption is 6.

It is well known that $T$ and $P$ have essential effects on hydrogen storage, where increasing $P$ and decreasing $T$ enhance the capacity of hydrogen storage. Thus, most studied systems are either under high $P$ or at very low $T$,[20] which may not be viable for mobile applications. For example, a storage capacity of 8 wt% for purified single wall carbon nanotubes (SWNTs) at 80 K with a hydrogen pressure of 13 MPa [32] and a lower hydrogen storage capacity of 2.3 wt% at 77 K were reported.[33] The hydrogen storage capacities in other carbon related materials, such as activated carbon (AC), single walled carbon nanohorn, SWNTs, and graphite nanofibers (GNFs) were also investigated.[34] Although the AC had a capacity of 5.7 wt% at 77 K with $P$ = 3 MPa, its capacity is < 1% at 300 K.[34] Recent experimental results demonstrated that the intrinsic graphene has hydrogen storage capacity of 1.7 wt% under 1 atm at 77 K, and 3 wt% under 100 atm at 298 K.[35] Thus, to meet the DOE target, it is necessary to study the adsorption and desorption behaviors of $H_2$ in the Al doped graphene at $T = 300$ K with different $P$. Therefore, the adsorption behaviors of $3H_2$/Al-doped-graphene and $6H_2$/Al-doped-graphene systems were calculated under 0.0001, 0.01, 0.1 and 1 GPa using *ab initio* MD simulation. For both the $3H_2$/Al-doped-graphene and $6H_2$/Al-doped-graphene systems, we found that all $H_2$ molecules were released at 0.0001 GPa [Fig. 5(c)]. However, there was only one $H_2$ molecule adsorbed in both the systems at 0.01 GPa, while the structure of the doped graphene was completely destroyed with H and Al forming covalent bond at 1 GPa [Fig. 5(d)]. When $P = 0.1$ GPa, there are three $H_2$ left on the top side of the two Al doped systems [Fig. 5(b)]. Therefore, the Al doped graphene for hydrogen storage capacity at room temperature and 0.1 GPa is 5.13 wt% with $E_b$ = -0.260 eV/$H_2$, satisfying the requirements of actual application. In addition, all the adsorbed $H_2$ molecules can be released when $P = 0.0001$ GPa.

## IV. CONCLUSION

In conclusion, the adsorption behaviors of $H_2$ in the intrinsic and Al doped graphene were studied using DFT. It is found that the physisorption of $H_2$ is greatly enhanced by doping Al into graphene. The doped Al varies the electronic structures of both C and $H_2$, causing the bands of $H_2$ overlapping with those of Al and C synchronously. It induces an intensive interaction between $H_2$ and the Al doped graphene. This was also demonstrated by the electron density distribution. In order to understand the temperature and pressure effects on the $H_2$ adsorption behavior for actual application, *ab initio* MD calculations for $H_2$/Al-doped-graphene system were processed. It is found that the system has 5.13 wt% hydrogen storage ability at $T$ = 300 K with $P$ = 0.1 GPa. Therefore, the Al doped graphene would be a promising hydrogen storage material owing to the strong interaction between $H_2$ and the Al doped graphene.

**ACKNOWLEDGMENTS**

This work was financially supported by National Key Basic Research and Development Program (Grant No. 2004CB619301), ″985 Project″ of Jilin University and Australia Research Council Discovery Program DP0665539.

Table 1. Summary of results for $H_2$ adsorption on intrinsic graphene and Al doped graphene on different adsorption sites. For $H_2$ adsorption on intrinsic graphene, there are 6 different adsorption sites as listed in the table. For $H_2$ adsorption on Al doped graphene, there are 18 different adsorption configurations as shown in Fig. 1.

| Initial binding configuration | | Intrinsic graphene | | Al doped graphene | | |
|---|---|---|---|---|---|---|
| | | $E_b$ (eV) | $d$ (Å)[b] | $E_b$ (eV) | $l$ (Å)[a] | $d$ (Å)[b] |
| $H_2$//graphene | T1 | -0.136 | 2.845 | -0.209 | 2.762 | |
| | T2 | | | -0.34 | 2.526 | 2.682 |
| | T3 | | | -0.407 | 2.588 | 2.486 |
| | T4 | | | -0.361 | 2.942 | 2.537 |
| | B1 | -0.139 | 2.817 | -0.21 | 2.757 | |
| | B2 | | | -0.411 | 2.527 | 2.575 |
| | B3 | | | -0.411 | 2.506 | 2.563 |
| | C1 | -0.159 | 2.635 | -0.427 | 2.083 | 2.073 |
| | C2 | | | -0.188 | | 2.657 |
| $H_2 \perp$ graphene | T1 | -0.141 | 2.615 | -0.153 | 2.622 | |
| | T2 | | | -0.284 | 2.427 | 2.749 |
| | T3 | | | -0.406 | 2.367 | 2.524 |
| | T4 | | | -0.33 | 2.976 | 2.179 |
| | B1 | -0.142 | 2.620 | -0.206 | 2.271 | 3.732 |
| | B2 | | | -0.412 | 2.468 | 2.595 |
| | B3 | | | -0.426 | 3.196 | 2.074 |
| | C1 | -0.148 | 2.425 | -0.426 | 2.092 | 2.104 |
| | C2 | | | -0.24 | 3.117 | 2.468 |

a. Distance between Al and $H_2$.
b. Distance between $H_2$ molecule and graphene or Al doped graphene layer.

Table 2. Charges of atoms in $H_2$ adsorbed in graphene system as well as charge transfer $Q$ between graphene and $H_2$ molecule, obtained by Mulliken analyse. The unit of the atom charge is one electron charge $e$, which is elided here for clarity.

| Atom | intrinsic graphene | Al doped graphene |
|------|--------------------|-------------------|
| Al1(C1) | 0.001 | 0.292 |
| C2 | -0.002 | -0.228 |
| C3 | 0 | -0.193 |
| C4 | 0 | -0.193 |
| H5 | -0.001 | -0.001 |
| H6 | -0.001 | 0.021 |
| $Q$ | -0.002 | 0.019 |

**Captions:**

FIG. 1. Nine different adsorption sites on Al doped graphene. The gray and pink balls are respectively C and Al atoms.

FIG. 2. The favorite adsorption configurations with 1 $H_2$ molecule adsorbed in intrinsic graphene (a), and in Al doped graphene (b). The gray and pink balls have the same meaning in Fig. 1, and the white balls are H atoms.

FIG. 3. The projected electronic density of states (PDOS) of adsorbed $H_2$, doped Al and graphene for both the $H_2$/graphene and $H_2$/Al-doped-graphene systems as shown in panel (a) and panel (b), respectively. Fermi level is set to 0.

FIG. 4. Electron density distributions in the $H_2$/graphene [panel (a)] and $H_2$/Al-doped-graphene [panel (b)] systems.

FIG. 5. Atomic configurations $3H_2$/Al-doped-graphene system at different temperature and pressure. (a) In the ideal condition with $T = 0$ K, (b) in the condition with $T = 300$ K and $P = 0.1$ GPa, (c) in the condition with $T = 300$ K and $P = 0.0001$ GPa, and (d) in the condition with $T = 300$ K and $P = 1$ GPa.

FIG. 1

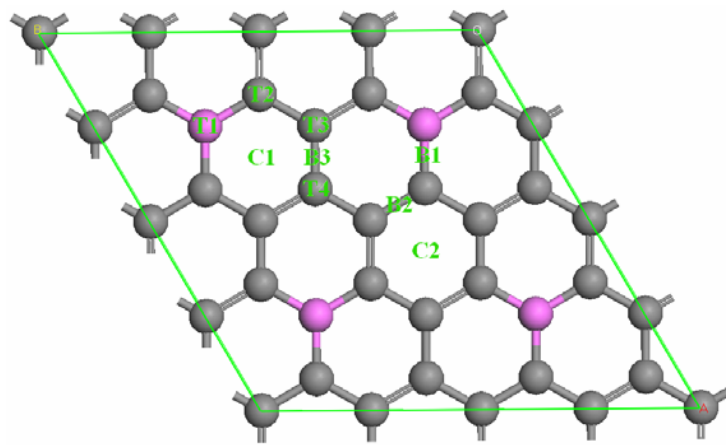

FIG. 2

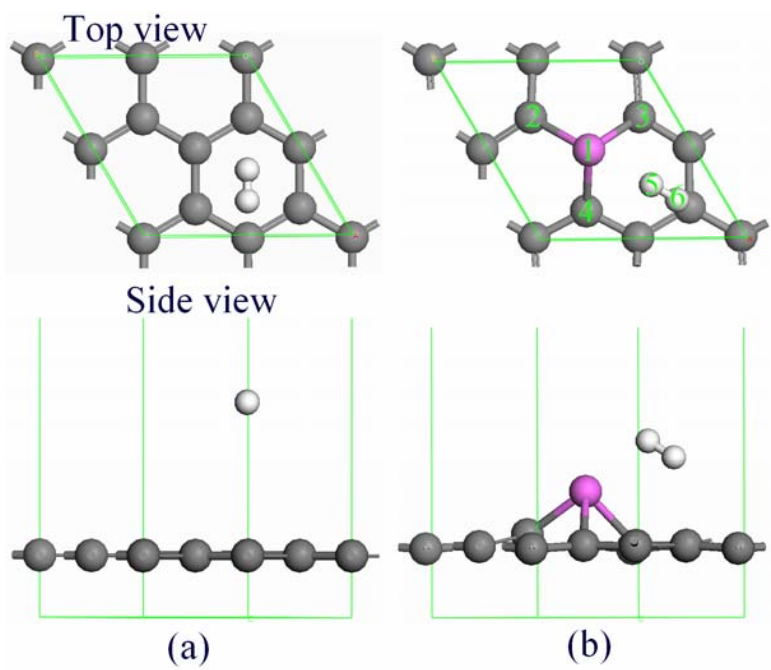

FIG. 3

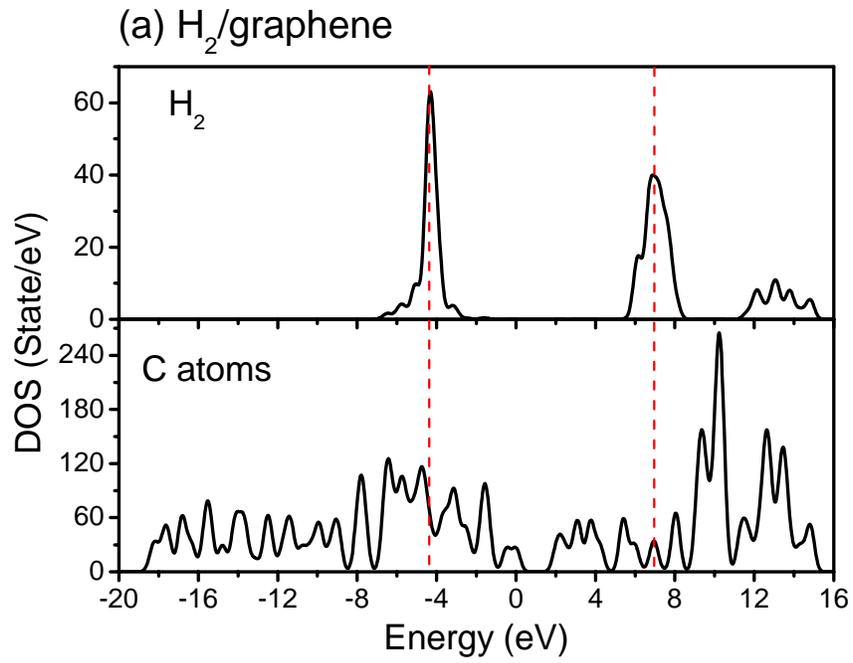

(a) H$_2$/graphene

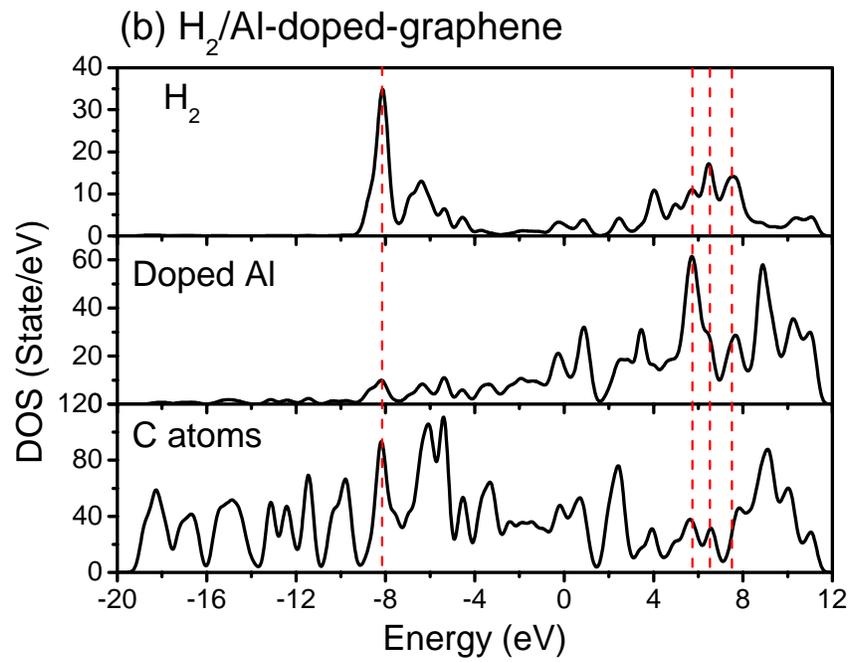

(b) H$_2$/Al-doped-graphene

FIG. 4

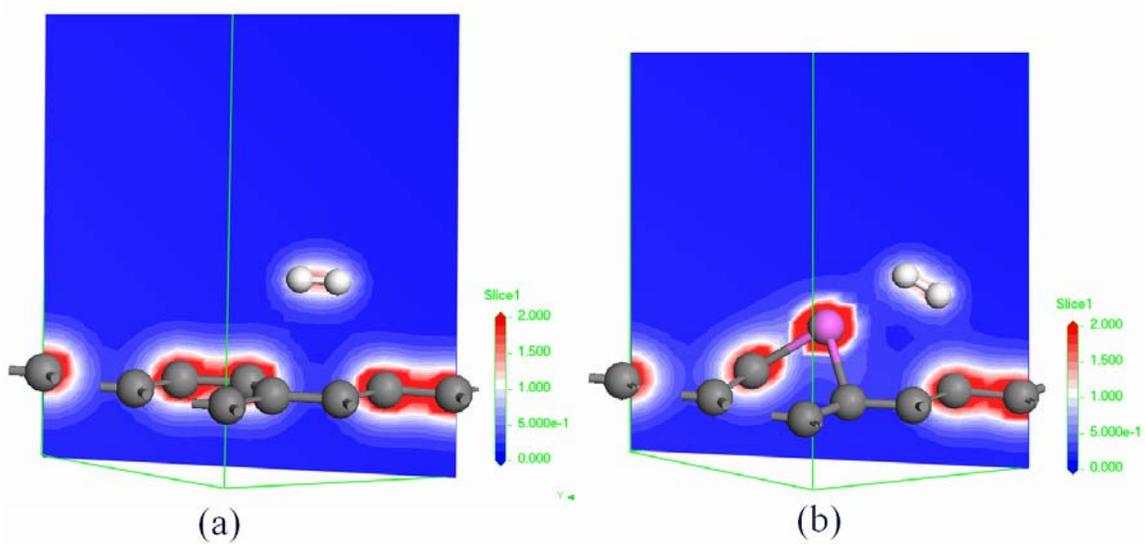

(a)                    (b)

FIG. 5

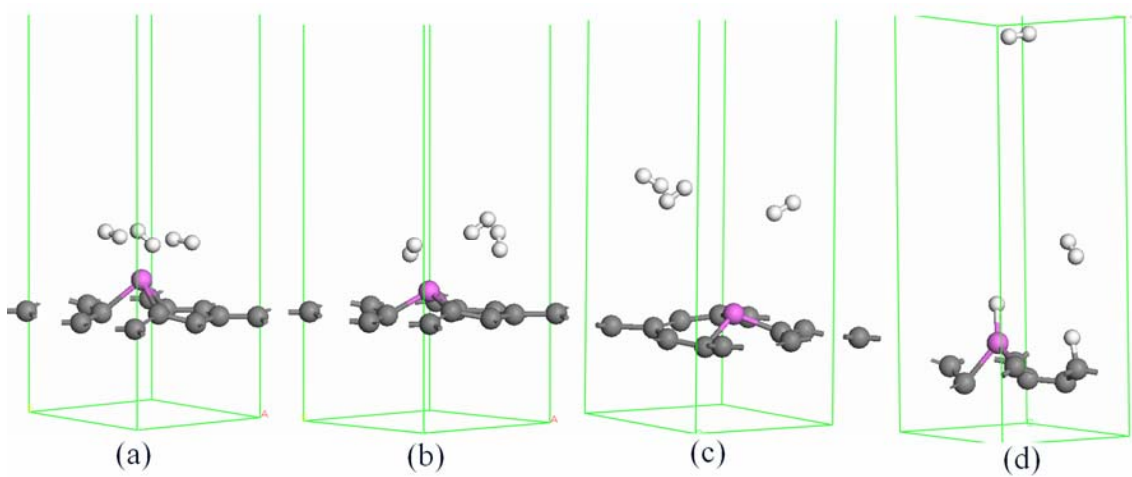

(a)        (b)        (c)        (d)